\documentclass[11pt,twoside]{./atmp}

\usepackage{amsmath,amssymb,youngtab}

\usepackage[all]{xy}
\usepackage{color}

\usepackage{amssymb,amsmath,dsfont}

\def\beq{\begin{equation}}
\def\eeq{\end{equation}}
\def\beqa{\begin{eqnarray}}
\def\eeqa{\end{eqnarray}}
\def\beqm{\begin{multline}}
\def\eeqm{\end{multline}}
\def\half{{\frac{1}{2}}}

\def\and{{\mathrm{and}}}
\def\Bid{{\mathds 1}}

\def\ket#1{|#1\rangle}

\def\CD{{\mathcal D}}
\def\CF{{\mathcal F}}

\def\CL{{\mathcal L}}
\def\CN{{\mathcal N}}

\def\BR{{\mathbb R}}

\def\BZ{{\mathbb Z}}

\DeclareMathOperator{\tr}{tr}
\DeclareMathOperator{\ind}{ind}
\DeclareMathOperator{\pexp}{Pexp}

\def\tyng#1{{\tiny\Yvcentermath1 \yng(#1)}}

\begin{document}

\title[Superconformal Index]
{Calculating the Superconformal Index and Seiberg Duality}

%\arxurl{DIAS-STP-07-13} %% arXiv number

\author[R\"omelsberger]{Christian R\"omelsberger}

\address{School of Theoretical Physics\\Dublin Institute for Advanced Studies\\10 Burlington Road\\Dublin 4\\Ireland}  %lines should be separated with double backslashes: \\
\addressemail{roemel@stp.dias.ie}

\begin{abstract}
We develop techniques to calculate an index for four dimensional superconformal field theories. This superconformal index is counting BPS operators which preserve only one supercharge. To calculate the superconformal index we quantize the field theory on $S^3\times\BR$ and show that the twisted theory has an appropriate mass gap. This allows for the interactions to be switched off continuously without the superconformal index being changed. We test those techniques for theories which go through a non-trivial RG flow and for Seiberg dual theories. This leads to the conjecture of some group/number theoretical identities.
\end{abstract}

\begin{flushright}
  DIAS-STP-07-13
\end{flushright}

\maketitle

\cutpage %move this line so that the first page breaks at the appropriate place.

\setcounter{page}{2}

\noindent

\section{Introduction}

Counting states in string and gauge theories has many applications:
\begin{itemize}
\item Knowing the spectrum of a theory is the first step towards understanding it and counting the states is a first step towards understanding the spectrum.
\item It allows to do thermodynamics and calculate something like (black hole) entropies.
\item It can be used as a check for dualities.
\end{itemize}
In general, however, it is very difficult to count the full spectrum of a given theory, but for supersymmetric theories it is often possible to count BPS states, which preserve a part of the supersymmetries. This still addresses the problems listed above.

Recently there has been a lot of progress in counting BPS states which preserve two supersymmetries in supersymmetric field theories \cite{Benvenuti:2006qr,Feng:2007ur,Hanany:2006uc,Forcella:2007wk,Forcella:2007ps,Mandal:2006tk,Dolan:2007rq}. Those BPS states are chiral operators and are related to holomorphic sections of the moduli space of vacua.

There also has been progress in understanding BPS states which just preserve one supersymmetry \cite{Romelsberger:2005eg,Kinney:2005ej,Nakayama:2005mf,Nakayama:2006ur,Nakayama:2007jy,Bianchi:2006ti}. Those gauge theory states are of great interest in string theory, because they could be microstates of AdS black holes which preserve just one supersymmetry \cite{Gutowski:2004ez,Kinney:2005ej}.

Most of the progress has been through an index which is a variation of the usual Witten index \cite{Witten:1982df} on $S^3\times\BR$ instead of $T^4$. This superconformal index has several chemical potentials for symmetry generators. Especially, there is one chemical potential $\mu$ for an energy operator $\Xi$ that resolves the degeneracy between an infinite number of BPS states.

The Lagrangian for $\CN=1$ supersymmetric field theories on $S^3\times\BR$ \cite{Romelsberger:2005eg} can be quantized. Because the twisted theory on $S^3\times\BR$ has a mass gap and the superconformal index is a topological quantity the interactions can be switched off continuously in order to calculate the superconformal index. The superconformal index only gets contributions from low lying states in the remaining infinite dimensional harmonic oscillator. This reduces calculating the superconformal index to a combinatoric and group theoretical problem.

This result can be used to calculate the first terms in a $e^{-\mu}$ expansion of the superconformal index for $SU(2)$ SQCD with three flavors and its Seiberg dual. Or other more complicated examples.

In section \ref{secindex} we review the definition of the superconformal index. In section \ref{secqm} we demonstrate the arguments for calculating the index on a quantum mechanics theory, which is a consistent truncation of the field theory. Later, in section \ref{indft} we generalize the arguments to field theories on $S^3\times\BR$ and propose a general formula (\ref{genletters}) and (\ref{pexpind}) for the superconformal index of a general gauge theory that flows to a conformal fixed point. Finally, in section \ref{secex} we calculate a couple of examples involving RG flows and Seiberg duality. This leads to the conjecture of some group/number theoretical identities (\ref{indeqsda}) and (\ref{indeqsdb}). 

\section{From the Witten index to the superconformal index}\label{secindex}

In most supersymmetric theories, one can define a Witten index. This is done by picking a supercharge $Q$ and its Hermitean conjugate $Q^\dagger$, they satisfy
\begin{align}
\{Q,Q\}&=0,\\
\{Q,Q^\dagger\}&=\Delta\qquad\and\label{qqdelta}\\
\{Q,(-1)^F\}&=0,
\end{align}
where in many cases $\Delta$ is the Hamiltonian.
Equation (\ref{qqdelta}) implies that the spectrum of $\Delta$ is bounded from below by zero. If $\Delta$ damps as well as the Hamiltonian $H$ of the theory
%, especially if it has a finite number of eigenstates with eigenvalue 0, then 
one can define the Witten index
\beq
\ind=\tr(-1)^F e^{-\beta\Delta}.
\eeq
This index is an integer, which is independent of $\beta$. However, if $\Delta$ has infinitely many eigenstates with zero eigenvalue, then the above expression is not well defined and one needs to find a better regulator $\Xi$, which commutes with $Q$
\beq
[\Xi,Q]=0\qquad\and\qquad\Xi\ge H
\eeq
to define the modified index
\beq
\ind(\mu)=\tr(-1)^Fe^{-\mu\Xi}.
\eeq
This modified index is actually a generating function for many indices. The coefficients of different powers of $t=e^{-\mu}$ are integers and topological invariants. This is similar to the elliptic genus or Gromov-Witten invariants.

Let us now turn to superconformal field theories. A superconformal field theory is naturally quantized on $S^3\times\BR$, this is called radial quantization. 
%For a superconformal theory one can define in such a way the superconformal index \cite{Romelsberger:2005eg,Kinney:2005ej}:  The symmetry group of $\CN=1$ superconformal field theories is $SU(2,2|1)$ which includes the conformal group $SO(2,4)$, the $U(1)$ R-symmetry group and eight supercharges. A superconformal theory can be mapped onto $S^3\times\BR$ by a conformal transformation. 
%
%Quantization with respect to the natural time on $S^3\times\BR$ is called radial quantization. 
The isometries of $S^3\times\BR$ are $SU(2)_L\times SU(2)_R\times\BR$ and the eight supercharges split up into a doublet $Q_\alpha$ under $SU(2)_L$ and a doublet $S_{\dot\alpha}$ under $SU(2)_R$. To define the superconformal index let us pick the supercharges $Q_1$ and $Q_1^\dagger$. They satisfy the anticommutation relation
\beq
\{Q_1,Q_1^\dagger\}=H-\frac{3}{2}R-2J_3=\Delta,\\
\eeq
where $H$ is the Hamiltonian in radial quantization, $R$ is the $U(1)$ R-charge and $J_3$ is a generator of $SU(2)_L$. It turns out that in a free theory $\Delta$ has infinitely many ground states, i.e. it does not damp well enough. However, there is an operator 
\beq
\Xi=H-\half R\ge\frac{2}{3}H
\eeq
which damps as well as the Hamiltonian\footnote{The inequality is due to a BPS bound implied by the anti-commutation relation $\{Q_\alpha,Q_\beta^\dagger\}=\delta_\alpha^\beta\left(H-\frac{3}{2}R\right)-4(\sigma^i)^\beta{}_\alpha J_i$.}. Furthermore, the $SU(2)_R$ and possibly an internal symmetry group\footnote{Those internal symmetries can be continuous or discrete.} $H$ commute with $Q_\alpha$
\begin{align}
[\gamma,Q_\alpha]&=0,\qquad\mathrm{for}\qquad \gamma\in SU(2)_R,\\
[h,Q_\alpha]&=0,\qquad\mathrm{for}\qquad h\in H.
\end{align}
The full superconformal index can be defined as
\beq
\ind(\mu,\gamma,h)=\tr (-1)^F e^{-\mu\Xi} \gamma h=\tr (-1)^F e^{-\mu(\Delta+R+2J^3)} \gamma h
\eeq
and only gets contributions from ground states of $\Delta$.

Note that the superconformal index is a group character of $\BR_\Xi\times SU(2)_R\times H$. This allows to split the superconformal index into representations of that group. In the examples involving Seiberg duality 
%\ref{secseiberg} 
we will do this.

The superconformal index is topological and is invariant under continuous deformations of the theory the following sense: A deformation of $Q_1$ and $\Delta$, that preserves $R+2J^3$, $\tilde J^i$ and the symmetry transformations $h\in H$ as well as the $SU(1|1)\times\BR_\Xi\times SU(2)_R\times H$ (super) symmetry group leaves the superconformal index invariant.

This also allows for non-conformal deformations. For this reason the superconformal index can be calculated in a weakly coupled UV theory that flows to a nontrivial conformal fixed point. In the next sections we will develop techniques for calculating the superconformal index for a given theory.

\section{The superconformal index of the Wess-Zumino quantum mechanics}\label{secqm}

As we will see in section \ref{indft} there is an infinite number of oscillator modes contributing to the superconformal index. This makes calculations fairly cumbersome. For this reason we will illustrate the main arguments of how to calculate the superconformal index in a related quantum mechanical theory, which is a consistent truncation of a Wess-Zumino model on $S^3\times\BR$ \cite{Romelsberger:2005eg}. In this quantum mechanical theory we will also be able to check those arguments by explicitely constructing the ground states.

\subsection{Canonical quatization}

The Lagrangian for a single chiral multiplet of R-charge $q$ is given by
\begin{multline}
L=\left(\partial_t-i\frac{3q-2}{2}\right)\phi^\dagger
\left(\partial_t+i\frac{3q-2}{2}\right)\phi-\phi^\dagger\phi-i\psi^\dagger\left(\partial_t+i\frac{3q-5}{2}\right)\psi\\
-|W^\prime|^2-\half W^{\prime\prime}\tilde\psi\psi-h.c.,
\end{multline}
where the superpotential is homogenous of R-charge 2, i.e.
\beq
W=\lambda\phi^n,
\eeq
with $nq=2$ and $\lambda$ a complex coupling constant.

Let us proceed in the canonical formalism. The canonical momenta are
\beq
\Pi=\dot\phi^\dagger-i\frac{3q-2}{2}\phi^\dagger,\qquad
\Pi^\dagger=\dot\phi+i\frac{3q-2}{2}\phi,\qquad\and\qquad
\pi^\alpha=-i\psi_\alpha^\dagger
\eeq
and satisfy the commutation relations
\beq
[\phi,\Pi]=-i\qquad
[\phi^\dagger,\Pi^\dagger]=-i\qquad\and\qquad
\{\psi_\alpha,\pi^\beta\}=-i\delta^\beta_\alpha.
\eeq
The Hamiltonian is
\begin{multline}
H=|\Pi|^2-i\frac{3q-2}{2}\Pi\phi+i\frac{3q-2}{2}\phi^\dagger\Pi^\dagger+|\phi|^2-\frac{3q-5}{2}\psi^\dagger\psi\\+|W^\prime|^2+W^{\prime\prime}\psi_1\psi_2-(W^{\prime\prime})^\dagger\psi_1^\dagger\psi_2^\dagger.
\end{multline}
Since $W$ has R-charge $2$, the theory is invariant under the $U(1)$ R-symmetry generated by
\beq
R=-iq(\Pi\phi-\phi^\dagger\Pi^\dagger)-(q-1)\psi^\dagger\psi,
\eeq
an $SU(2)_L$ symmetry that acts on the spinors, which is generated by
\beq
J^i=-\psi^\dagger_\alpha\sigma^i{}_\alpha{}^\beta\psi_\beta
\eeq
and the supersymmetries generated by
\beq
Q_\alpha=(\Pi-i\phi^\dagger)\psi_\alpha+i(W^\prime)^\dagger\epsilon_{\alpha\beta}\psi_\beta^\dagger.
\eeq
This symmetry group is $SU(2|1)$
\beq
\{Q_\alpha,Q_\beta\}=0,\quad
\{Q_\alpha^\dagger,Q_\beta^\dagger\}=0\quad\and\quad
\{Q_\alpha,Q_\beta^\dagger\}=\delta_\alpha^\beta\,(H-\frac{3}{2}R)-4\sigma^i{}_\alpha{}^\beta J^i.
\eeq

We now pick the supercharge $Q_1$, which satisfies the relations
\begin{align}
\{Q_1,Q_1^\dagger\}&=\Delta\equiv H-\frac{3}{2}R-2J^3\qquad\and\\
[\Xi,Q_1&]=0,\qquad\mathrm{where}\qquad\Xi=H-\half R
\end{align}
and define the superconformal index
\beq
\ind(\mu)=\tr(-1)^Fe^{-\mu\Xi}=\tr(-1)^F t^{\Delta+R+2J^3},
\eeq
where  $t=e^{-\mu}$.

\subsection{The free theory}

In the next subsection we will show that the superconformal index of the interacting theory can be reduced to the superconformal index of the free theory. For this reason let us start by calculating the superconformal index of the free theory. Let us formulate the theory in terms of the annihilation and creation operators
\begin{align}
a_1&=\frac{1}{\sqrt{2}}(\Pi^\dagger+i\phi),&
a_1^\dagger&=\frac{1}{\sqrt{2}}(\Pi-i\phi^\dagger),\\
a_2&=\frac{1}{\sqrt{2}}(\Pi+i\phi^\dagger),&
a_2^\dagger&=\frac{1}{\sqrt{2}}(\Pi^\dagger-i\phi),\\
b_1&=\psi_1,&
b_1^\dagger&=\psi_1^\dagger,\\
b_2&=\psi_2,&
b_2^\dagger&=\psi_2^\dagger.
\end{align}
The twisted Hamiltonian of the free theory is then
\beq
H-\frac{3}{2}R-2J^3=2a_1^\dagger a_1+2b_1^\dagger b_1.
\eeq
It's ground states are 
\beq
\ket{0,m,0,\eta}.
\eeq
and the commuting $U(1)$ is
\beq
R+2J^3=-qa_1^\dagger a_1+qa_2^\dagger a_2-qb_1^\dagger b_1-(q-2)b_2^\dagger b_2.
\eeq
This implies that the superconformal index is
\beq\label{eqindfree}
\ind(\mu)=\frac{1-t^{2-q}}{1-t^q}=\sum_{m=0}^{n-2}t^{qm},
\eeq
where $qn=2$. This finite result comes from a cancellation between fermionic and bosonic states with $R+2J^3\ge(n-1)q$. To avoid this cancellation in the free theory, one can introduce a chemical potential for an additional internal $U(1)$ symmetry generated by
\beq
T=-a_1^\dagger a_1+a_2^\dagger a_2-b_1^\dagger b_1-b_2^\dagger b_2.
\eeq
The internal $U(1)$ symmetry commutes with the supercharges and an appropriate chemical potential can be inserted into the superconformal index
\beq
\ind(\mu,u)=\tr(-1)^Ft^{\Delta+R+2J^3}u^T=\frac{1-t^{(n-1)q}u^{-1}}{1-t^qu}.
\eeq
This resolves all the degeneracies in the superconformal index for the free theory.

\subsection{The interacting theory}\label{ssecwz}

In the interacting theory the extra $U(1)$ symmetry is broken by the superpotential and one expects that the canceling states in (\ref{eqindfree}) are actually lifted by the interaction. We now use two independent arguments to support this claim:

The superconformal index 
\beq
\ind(\mu)=\tr(-1)^Ft^{\Xi}=\tr(-1)^Ft^{\Delta+R+2J^3}
\eeq
is topological and does not change under continuous deformations of $Q_1$ and $\Delta$ as long as $R+2J^3$, $\tilde J^i$ and $h$ remain invariant and the (super) symmetry algebra is preserved. For this reason we can switch off the superpotential as long as we can argue that the eigenstates of $\Xi=\Delta+R+2J^3$ remain sufficiently localized around the origin in field space. Indeed $\Xi$ has a mass gap and localizes its eigenstates, even though $\Delta$ has the zero modes $a_2$ and $b_2$.

For this reason the superpotential can be switched off continuously and and the superconformal index can be calculated by counting excitations of free bosonic and fermionic oscillators. The only manifestation of the superpotential is due to the fact that the extra $U(1)$ symmetry is broken and the cancellations of the previous section can happen, leading to
\beq
\ind(\mu)=\sum_{m=0}^{n-2}t^{qm}.
\eeq

Using a 'zig-zag' argument similar to \cite{Kachru:1993pg,Bott} we can actually verify this result for the superconformal index and show, that in the interacting theory all the canceling states of the free theory calculation are actually lifted. The ground states of the twisted Hamiltonian $\Delta$ are in one to one correspondence with the cohomology of the supercharge $Q_1^\dagger$. 

The supercharge $Q_1^\dagger$ is the sum 
\beq
Q_1^\dagger=\sqrt{2}a_1b_1^\dagger-i\lambda\left(\frac{a_1-a_2^\dagger}{\sqrt{2}i}\right)^{n-1}b_2=Q_1^{(0)}{}^\dagger+\lambda Q_1^{(1)}{}^\dagger
\eeq
of the supercharge of the free theory and an interaction term. We now follow a perturbative procedure and expand a ground state as
\beq
\ket{\Omega}=\ket{\Omega^{(0)}}+\lambda\ket{\Omega^{(1)}}+\lambda^2\ket{\Omega^{(2)}}+\cdots,
\eeq
where all the terms $\ket{\Omega^{(l)}}$ have the same $R+2J^3$ charge. Then the action of $Q_1^\dagger$ on $\ket{\Omega}$ can be expanded as follows:
\beq\begin{split}
Q_1^\dagger\ket{\Omega}=Q_1^{(0)}{}^\dagger\ket{\Omega^{(0)}}&+\lambda\left(Q_1^{(0)}{}^\dagger\ket{\Omega^{(1)}}+Q_1^{(1)}{}^\dagger\ket{\Omega^{(0)}}\right)\\
&+\lambda^2\left(Q_1^{(0)}{}^\dagger\ket{\Omega^{(2)}}+Q_1^{(1)}{}^\dagger\ket{\Omega^{(1)}}\right)+\cdots
\end{split}\eeq
For a state to be closed, it needs to satisfy
\beq\label{pertclosed}
Q_1^{(0)}{}^\dagger\ket{\Omega^{(0)}}=0\quad\and\quad
Q_1^{(0)}{}^\dagger\ket{\Omega^{(l+1)}}+Q_1^{(1)}{}^\dagger\ket{\Omega^{(l)}}=0
\eeq
and for a state to be exact it needs to satisfy
\beq\label{pertexact}
%\ket{\Omega^{(0)}}=Q_1^{(0)}{}^\dagger\ket{\Lambda^{(0)}}\quad\and\quad
\ket{\Omega^{(l)}}=Q_1^{(0)}{}^\dagger\ket{\Lambda^{(l)}}+Q_1^{(1)}{}^\dagger\ket{\Lambda^{(l-1)}}.
\eeq
In the example at hand this expansion actually terminates after the first step.

At leading order, we have to find the cohomology of $Q_1^{(0)}{}^\dagger$. As we have seen before, the cohomology of $Q_1^{(0)}{}^\dagger$ can be represented by $\ket{0,m,0,\eta}$, the ground states of the free Hamiltonian. The  $Q_1^{(0)}{}^\dagger$-exact states have the form $\ket{m_1,m_2,1,\eta}$. 

Next, equation (\ref{pertclosed}) implies that $\ket{\Omega^{(0)}}$ needs to be $Q_1^{(1)}{}^\dagger$-closed inside the $Q_1^{(0)}{}^\dagger$-cohomology 
\beq
Q_1^{(1)}{}^\dagger\ket{\Omega^{(0)}}=-Q_1^{(0)}{}^\dagger\ket{\Omega^{(1)}}
\eeq
which only has a solution for $\eta=0$, where it reduces to
 \beq
 Q_1^{(0)}{}^\dagger\ket{\Omega^{(1)}}=0,
 \eeq
 i.e. $\ket{\Omega^{(1)}}$ has to be again in the $Q_1^{(0)}{}^\dagger$-cohomology. Furthermore, the  relation (\ref{pertexact}) implies that states of the form
 \beq
 \ket{\Omega^{(0)}}\sim\ket{0,m+n-1,0,0}
 \eeq
 are exact. One can now iterate the procedure for $\ket{\Omega^{(1)}}$ and get the same result. For this reason one can just choose 
 \beq
 \ket{\Omega^{(l)}}=0\qquad{\mathrm{for}}\qquad l\ge 1
 \eeq
 and the expansion terminates after the first step.This shows that the cohomology of $Q_1^\dagger$ can be represented by the states
\beq
\ket{0,m,0,0},\qquad m=0,\cdots,n-2
\eeq
proving the claim before.

The Wess-Zumino model actually also has a discrete $\BZ_n$ symmetry left of the $U(1)$ in the free theory and one could define the superconformal index
\beq
\ind(\mu,\omega)=\tr(-1)^F t^\Xi g=\sum_{m=0}^{n-2}t^{qm}\omega^m,
\eeq
where $g\in\BZ_n$ and $\omega^n=1$. However this does not lead to any new information.

\section{The superconformal index for field theories}\label{indft}

In this section we will generalize the quantum mechanical result of the previous section to field theory. In \cite{Romelsberger:2005eg} the Lagrangian for $\CN=1$ field theory on $S^3\times\BR$ was derived. For a Wess-Zumino model of a chiral multiplet with R-charge $q$ the Lagrangian is
\begin{align}
\CL_0&=\left(\partial_t-i\frac{3q-2}{2}\right)\phi^\ast
\left(\partial_t+i\frac{3iq-2}{2}\right)\phi-
4\sigma^{(L)}_i\phi^\ast\sigma^{(L)}_i\phi-\phi^\ast\phi\\\nonumber
&+i\bar\psi\gamma^0\left(\partial_t+i\frac{3q-2}{2}\right)\psi-
2i\bar\psi\gamma^i\left(\sigma^{(L)}_i+\frac{1}{8}\epsilon_{ijk}\gamma^{jk}\right)\psi+F^\ast F,\\
\CL_W&=W^\prime(\phi)F-\half W^{\prime\prime}(\phi)\tilde\psi\psi+{\rm h.c.}
\end{align}
with
\beq
W^\prime(\phi)=c\phi^{-\frac{q-2}{q}},
\eeq
where $\sigma^{(L)}_i$ are the derivatives with respect to the Killing vector fields generating $SU(2)_L$ and the spinors are represented in a vielbein basis given by the right invariant 1-forms. The Lagrangian for a gauge theory is given in appendix \ref{applagr}. In general the superconformal R-charges $q$ have to be fixed by a-maximization \cite{Intriligator:2003jj}.

As we did for the quantum mechanical theory, let us first understand the free field theory.

\subsection{The free chiral multipet}

The Lagrangian for the boson in a free chiral multiplet of R-charge $q$ is
\beq
\CL_{\phi}=(\partial_t-i\frac{3q-2}{2})\phi^\dagger(\partial_t+i\frac{3q-2}{2})\phi-4\sigma^{(L)}_i\phi^\dagger\sigma^{(L)}_i\phi-\phi^\dagger\phi.
\eeq
When the scalar field is decomposed into spherical harmonics
\beq
\phi=\sum_{j,j_3,\tilde j_3}\phi_{j,j_3,\tilde j_3}Y_{j,j_3,\tilde j_3}.
\eeq
the Lagrangian has the form
\begin{multline}
L_{\phi}=\\
\sum_{j,j_3,\tilde j_3}\left((\partial_t-i\frac{3q-2}{2})\phi_{j,j_3,\tilde j_3}^\dagger(\partial_t+i\frac{3q-2}{2})\phi_{j,j_3,\tilde j_3}-(2j+1)^2\phi_{j,j_3,\tilde j_3}^\dagger\phi_{j,j_3,\tilde j_3}\right).
\end{multline}
The canonical momenta are
\beq
\Pi_{j,j_3,\tilde j_3}=(\partial_t-i\frac{3q-2}{2})\phi_{j,j_3,\tilde j_3}^\dagger\quad\and\quad
\Pi_{j,j_3,\tilde j_3}^\dagger=(\partial_t+i\frac{3q-2}{2})\phi_{j,j_3,\tilde j_3},
\eeq
the Hamiltonian is
\begin{multline}
H_{\phi}=\sum_{j,j_3,\tilde j_3}\Big(\Pi_{j,j_3,\tilde j_3}\Pi_{j,j_3,\tilde j_3}^\dagger-i\frac{3q-2}{2}\Pi_{j,j_3,\tilde j_3}\phi_{j,j_3,\tilde j_3}+i\frac{3q-2}{2}\phi_{j,j_3,\tilde j_3}^\dagger\Pi_{j,j_3,\tilde j_3}^\dagger+\\
(2j+1)^2\phi_{j,j_3,\tilde j_3}^\dagger\phi_{j,j_3,\tilde j_3}\Big),
\end{multline}
the R-charge is
\beq
R_{\phi}=-iq\sum_{j,j_3,\tilde j_3}\left(\Pi_{j,j_3,\tilde j_3}\phi_{j,j_3,\tilde j_3}-\phi_{j,j_3,\tilde j_3}^\dagger\Pi_{j,j_3,\tilde j_3}^\dagger\right),
\eeq
the $J^3$ charge is
\beq
J^3_{\phi}=-i\sum_{j,j_3,\tilde j_3}j_3\left(\Pi_{j,j_3,\tilde j_3}\phi_{j,j_3,\tilde j_3}-\phi_{j,j_3,\tilde j_3}^\dagger\Pi_{j,j_3,\tilde j_3}^\dagger\right)
\eeq
and 
the $\tilde J^3$ charge is
\beq
\tilde J^3_{\phi}=-i\sum_{j,j_3,\tilde j_3}\tilde j_3\left(\Pi_{j,j_3,\tilde j_3}\phi_{j,j_3,\tilde j_3}-\phi_{j,j_3,\tilde j_3}^\dagger\Pi_{j,j_3,\tilde j_3}^\dagger\right).
\eeq
The twisted Hamiltonian is then
\begin{multline}
\Delta_{\phi}=H_{\phi}-\frac{3}{2}R_{\phi}-2J^3_{\phi}=\\
%\sum_{j,j_3,\tilde j_3}\Big(\Pi_{j,j_3,\tilde j_3}\Pi_{j,j_3,\tilde j_3}^\dagger+\\i(2j_3+1)\Pi_{j,j_3,\tilde j_3}\phi_{j,j_3,\tilde j_3}-i(2j_3+1)\phi_{j,j_3,\tilde j_3}^\dagger\Pi_{j,j_3,\tilde j_3}^\dagger+(2j+1)^2\phi_{j,j_3,\tilde j_3}^\dagger\phi_{j,j_3,\tilde j_3}\Big)=\\
2\sum_{j,j_3,\tilde j_3}\Big((j+j_3+1)a_{1,j,j_3,\tilde j_3}^\dagger a_{1,j,j_3,\tilde j_3}+(j-j_3)a_{2,j,j_3,\tilde j_3}^\dagger a_{2,j,j_3,\tilde j_3}\Big),
\end{multline}
with the ladder (annihilation) operators
\begin{align}
a_{1,j,j_3,\tilde j_3}&=\frac{1}{\sqrt{4j+2}}\big(\Pi_{j,j_3,\tilde j_3}^\dagger+i(2j+1)\phi\big)\qquad\and\\
a_{2,j,j_3,\tilde j_3}&=\frac{1}{\sqrt{4j+2}}\big(\Pi_{j,j_3,\tilde j_3}+i(2j+1)\phi^\dagger\big).
\end{align}
The zero modes are given by
\beq
a_{2,j,j,\tilde j_3}\qquad\and\qquad a_{2,j,j,\tilde j_3}^\dagger
\eeq
and the $R+2J^3$ charge is
\beq
R_{\phi}+2J^3_{\phi}=-\sum_{j,j_3,\tilde j_3}(q+2j_3)\Big(a_{1,j,j_3,\tilde j_3}^\dagger a_{1,j,j_3,\tilde j_3}-a_{2,j,j_3,\tilde j_3}^\dagger a_{2,j,j_3,\tilde j_3}\Big).
\eeq

As in the quantum mechanical theory, the theory has an extra $U(1)$ symmetry generated by
\beq
T_{\phi}=-\sum_{j,j_3,\tilde j_3}\Big(a_{1,j,j_3,\tilde j_3}^\dagger a_{1,j,j_3,\tilde j_3}-a_{2,j,j_3,\tilde j_3}^\dagger a_{2,j,j_3,\tilde j_3}\Big).
\eeq
The generating function for all the oscillator modes contributing to the superconformal index is
\beq
f_{\phi}(t,y,u)=\frac{t^{q}u}{(1-ty)(1-\frac{t}{y})},
\eeq
where $t$ is the fugacity for the $R+2J^3$-charge, $u$ is the fugacity for the $T$-charge and $y$ is the fugacity for the $\tilde J^3$-charge. 

Similarly, the Lagrangian for the fermion in a free chiral multiplet of R-charge $q$ is
\beq
\CL_\psi=i\bar\psi\gamma^0\left(\partial_0+i\frac{3q-2}{2}\right)\psi-
2i\bar\psi\gamma^i\left(\sigma^{(L)}_i+\frac{1}{8}\epsilon_{ijk}\gamma^{jk}\right)\psi.
\eeq
A similar expansion in spherical harmonics leads to a generating function
\beq
f_{\psi}(t,y,u)=-\frac{t^{2-q}u^{-1}}{(1-ty)(1-\frac{t}{y})}
\eeq
for the fermionic oscillator modes contributing to the superconformal index leading to the generating function
\beq
f_{\Phi}(t,y,u)=\frac{t^{q}u-t^{2-q}u^{-1}}{(1-ty)(1-\frac{t}{y})}
\eeq
for the oscillator modes of a free chiral multiplet contributing to the superconformal index.

To get the superconformal index one has to count all the oscillator excitations, which is done by the Plethystic exponential \cite{Aharony:2003sx,Kinney:2005ej,Benvenuti:2006qr,Feng:2007ur}
\beq
\ind(\mu,y,u)=g_{\Phi}(t,y,u)=\pexp(f_{\Phi}(t,y,u))=\exp\left(\sum_{l=1}^{\infty}\frac{1}{l}f_{\Phi}(t^l,y^l,u^l)\right),
\eeq
which in this case is
\beq
\ind(\mu,y,u)=\prod_{m,n}\frac{1-t^{2-q+m+n}y^{m-n}u^{-1}}{1-t^{q+m+n}y^{m-n}u}.
\eeq

\subsection{Wess-Zumino models}\label{WZfieldtheory}

Let us now consider an ungauged theory of chiral multiplets with interactions given by a homogenous superpotential. Similar as in the quantum mechanical theory in section \ref{ssecwz}, there is a mass gap due to the R-charge dependent 'curvature couplings' in the Lagrangian. For this reason, the superpotential can actually be switched off for the purpose of calculating the superconformal index.

However, the superpotential breaks the group $G$ of internal symmetries of the free theory to a subgroup $H$ and only this subgroup can be used to resolve degeneracies in the superconformal index. If the interacting theory has chiral multiplets $\Phi_i$ of R-charge $q_i$ transforming in the representation $R_i$ of the global symmetry group $H$, then the generating function for the oscillator modes contributing to the superconformal index is
\beq
f_{\Phi}(t,\gamma,h)=\sum_i\frac{t^{q_i}\chi_{R_i}(h)-t^{2-q_i}\chi_{\bar R_i}(h)}{1-t\chi_f(\gamma)+t^2}
\eeq
and the superconformal index is the Plethystic exponential thereof
\beq
\ind(\mu,\gamma,h)=%g_{\Phi}(t,y,h)=
\exp\left(\sum_{l=1}^{\infty}\frac{1}{l}f_{\Phi}(t^l,\gamma^l,h^l)\right).
\eeq
Note that the superconformal index can again be viewed as a character of the $\BR_\Xi\times SU(2)_R\times H$ group. This allows to extract contributions transforming in given irreducible representations.

For a Wess-Zumino model of a single chiral multiplet with a superpotential
\beq
W(\phi)=\phi^n,
\eeq
the global $U(1)$-symmetry is broken to $\BZ_n$ and the superconformal index is generated by
\beq
f_{\Phi}(t,\gamma,\omega)=\frac{t^{q}\omega-t^{2-q}\omega^{-1}}{1-t\chi_f(\gamma)+t^2}=\frac{t^{q}\omega-(t^{q}\omega)^{n-1}}{(1-ty)(1-\frac{t}{y})},
\eeq
where $nq=2$ and $\omega^n=1$. One can extract the lowest contributions, which are singlets under the $SU(2)_R$
\beq
\ind(\mu,\gamma,\omega)=1+t^{q}\omega+\cdots+(t^{q}\omega)^{n-2}+\cdots
\eeq
Those contributions agree with the superconformal index of the quantum mechanical theory and they are due to the chiral primaries.

If the superpotential includes a mass term for a field, say
\beq
W(\phi)=\phi^2
\eeq
then this superfield should be integrated out in the IR, i.e. just disappear from the low energy theory. For this reason it should not contribute to the superconformal index. Indeed, such a superfield has R-charge $1$, the global symmetry is broken to $\BZ_2$ or nothing and the superfield does not contribute to the superconformal index
\beq
f_{\Phi}(t,\gamma,\omega)=\frac{t\omega-t\omega^{-1}}{1-t\chi_f(\gamma)+t^2}=0.
\eeq

\subsection{Gauge theories}

A free gauge theory can be treated in a similar way as the free chiral multiplet except for a few modifications related to gauge fixing and the Gauss law constraint. The most convenient gauge in the case at hand is the temporal gauge $A_0=0$. In the free limit, the dynamics of the gauge theory is described by free abelian gauge fields together with an overall Gauss law constraint imposing the condition, that all states are gauge singlets.

The generating function for oscillator modes of a free abelian vector multiplet contributing to the superconformal index is
\beq
f_V(\mu,\gamma)=\frac{2t^2-t\chi_f(\gamma)}{1-t\chi_f(\gamma)+t^2}.
\eeq
For a nonabelian gauge theory with gauge group $G$ one has to substitute the group character in the adjoint representation
\beq
f_V(\mu,\gamma,g)=\frac{2t^2-t\chi_f(\gamma)}{1-t\chi_f(\gamma)+t^2}\,\chi_{adj}(g).
\eeq
The matter fields in representations $r_i$ of the gauge group $G$ lead to contributions of the form
\beq
f_{\Phi}(t,\gamma,h,g)=\sum_i\frac{t^{q_i}\chi_{R_i}(h)\chi_{r_i}(g)-t^{2-q_i}\chi_{\bar R_i}(h)\chi_{\bar r_i}(g)}{1-t\chi_f(\gamma)+t^2}
\eeq
so that the total generating function for the oscillator modes is
\begin{multline}\label{genletters}
f(t,\gamma,h,g)=\\
\frac{2t^2-t\chi_f(\gamma)}{1-t\chi_f(\gamma)+t^2}\,\chi_{adj}(g)+\sum_i\frac{t^{q_i}\chi_{R_i}(h)\chi_{r_i}(g)-t^{2-q_i}\chi_{\bar R_i}(h)\chi_{\bar r_i}(g)}{1-t\chi_f(\gamma)+t^2}.
\end{multline}

The superconformal index is then given by the gauge invariant terms in the Plethystic exponential
\beq\label{pexpind}
\ind(\mu,\gamma,h)=\int_G [dg]\,\exp\left(\sum_{l=1}^{\infty}\frac{1}{l}f(t^l,\gamma^l,h^l,g^l)\right).
\eeq

As for ungauged theories the gauge theories have a mass gap and the arguments of section \ref{ssecwz} go through. For this reason the superconformal index can be calculated in the free theory. The interactions  manifest themselves in the preserved global symmetry group $H$. This agrees with the fact that the superconformal index is a topological quantity and should only depend on the universality class of the theory. Here the universality class of the theory is just given by the matter content and the symmetry group of the theory. All interactions that are consistent with the symmetry group are allowed.

\section{Examples}\label{secex}

Let us now calculate the superconformal index for a couple examples.

\subsection{Leigh-Strassler flows}

First let us look at the  renormalization group flow of $\CN=4$ super Yang-Mills theory that is deformed by a mass term for one of the adjoint chiral multiplets. This flow was discussed by Leigh and Strassler \cite{Leigh:1995ep} and in the context of holographic renormalization group flows by Freedman, Gubser, Pilch and Warner \cite{Freedman:1999gp,Pilch:2000fu}. In this theory the superpotential for the three adjoint chiral multiplets $X$, $Y$ and $Z$ is deformed by a mass term for $Z$
\beq
W=\tr [X,Y]Z \mapsto W=\tr [X,Y]Z + \frac{m}{2}\tr Z^2.
\eeq
This breaks the $SO(6)$ R-symmetry to a global $SU(2)$ symmetry and a $U(1)_R$ symmetry. The charges of the chiral multiplets are\\
\begin{center}
\begin{tabular}{|c|c|c|}
\hline
Field & $SU(2)$ & $U(1)_R$ \\
\hline
$(X,Y)$ & $2$ & $\half$ \\
$Z$ & $1$ & $1$ \\
\hline
\end{tabular}
\end{center}

As already mentioned in section \ref{WZfieldtheory} the scalar $Z$ has R-charge $1$ and does not contribute to the superconformal index. This obviously agrees with the theory in the infrared, where $Z$ has been integrated out. The same arguments apply for the Klebanov-Witten flow \cite{Klebanov:1998hh,Halmagyi:2004jy,Halmagyi:2005pn} and related flows.

\subsection{Seiberg duality for a $SU(2)$ super Yang-Mills theory with three flavors}\label{secseiberga}

Let us now turn to a more interesting example and compare the superconformal index of two theories related by a strong coupling duality, 
%The superconformal index is a nontrivial check for nonperturbative dualities of superconformal field theories. 
Seiberg duality of SQCD in the conformal window $\frac{3}{2}N_c\le N_f\le 3N_c$. Seiberg duality \cite{Seiberg:1994pq} is the duality in the IR between an 'electric' $SU(N_c)$ gauge theory with $N_f$ flavors and a 'magnetic' $SU(N_f-N_c)$ gauge theory with $N_f$ flavors.
%, a gauge invariant, elementary 'Meson' field together with a cubic superpotential. 

For simplicity let us first look at $SU(2)$ SQCD with 3 flavors. This theory has actually 6 quark fields transforming under a global $SU(6)$ flavor symmetry group. The oscillators contributing to the superconformal index are summarized in the following table
\begin{center}
\begin{tabular}{|l|c|c|c|c|c|}
\hline
Theory & Field & $SU(2)_c$ & $SU(6)$ & $SU(2)_R$ & $3\Xi$ \\\hline
electric & $\phi^{(l)}$ & $2$ & $6$ & $l+1$ & $1+3l$ \\
& $\psi^{(l)}$ & $2$ & $\bar 6$ & $l+1$ & $5+3l$ \\
& $\lambda^{(l)}$ & $3$ & $1$ & $l\oplus(l+2)$ & $3+3l$ \\
& $F^{(l)}$ & $3$ & $1$ & $(l+1)\oplus(l+1)$ & $6+3l$ \\\hline
magnetic & $\phi^{(l)}_M$ & N/A & $15$ & $l+1$ & $2+3l$ \\ 
& $\psi^{(l)}_M$ & N/A & $\overline{15}$ & $l+1$ & $4+3l$ \\\hline
\end{tabular}
\end{center}
where $l$ labels the appropriate harmonic on $S^3$.

The first terms in the $t$-expansion of the superconformal index of the electric theory are summarized in the following table
\begin{center}
\begin{tabular}{|c|l|}
\hline
$3\Xi$ &  $SU(6)\times SU(2)_R$ \\\hline
$0$ & $(1,1)$ \\\hline
$2$ & $\underbrace{(\tyng{1,1},1)}_{(\phi^{(0)})^2}$ \\\hline
$4$ & $\underbrace{(\tyng{2,2},1)}_{(\phi^{(0)})^4}$ \\\hline
$5$ & $\underbrace{(\tyng{2},\tyng{1})\oplus(\tyng{1,1},\tyng{1})}_{\phi^{(0)}\phi^{(1)}}\ominus\underbrace{(\tyng{2},\tyng{1})}_{(\phi^{(0)})^2\lambda^{(0)}}$ \\\hline
$6$ & $\underbrace{(\tyng{3,3},1)}_{(\phi^{(0)})^6}\oplus\underbrace{(1,1)}_{(\lambda^{(0)})^2}\ominus\underbrace{(1,1)\ominus(\tyng{2,1,1,1,1},1)}_{\phi^{(0)}\psi^{(0)}}$ \\\hline
$7$ & $\underbrace{(\tyng{3,1},\tyng{1})\oplus(\tyng{2,2},\tyng{1})\oplus(\tyng{2,1,1},\tyng{1})}_{(\phi^{(0)})^3\phi^{(1)}}\ominus\underbrace{(\tyng{3,1},\tyng{1})}_{(\phi^{(0)})^4\lambda^{(0)}}$ \\\hline
$8$ & $\underbrace{(\tyng{2},1)\oplus(\tyng{2},1)}_{F^{(0)}(\phi^{(0)})^2}\oplus\underbrace{(\tyng{2},\tyng{2})\oplus(\tyng{1,1},1)}_{(\lambda^{(0)})^2(\phi^{(0)})^2}\ominus\underbrace{(\tyng{2},1)\ominus(\tyng{2},\tyng{2})}_{\lambda^{(1)}(\phi^{(0)})^2}$ \\
& $\oplus\underbrace{(\tyng{4,4},1)}_{(\phi^{(0)})^8}\ominus\underbrace{(\tyng{2},\tyng{2})\ominus(\tyng{2},1)\ominus(\tyng{1,1},\tyng{2})\ominus(\tyng{1,1},1)}_{\phi^{(0)}\phi^{(1)}\lambda^{(0)}}\oplus
\underbrace{(\tyng{1,1},\tyng{2})\oplus(\tyng{2},1)}_{(\phi^{(1)})^2}$ \\
& $\oplus\underbrace{(\tyng{2},\tyng{2})\oplus(\tyng{1,1},\tyng{2})}_{\phi^{(0)}\phi^{(2)}}\ominus\underbrace{(\tyng{3,2,1,1,1},1)\ominus(\tyng{2},1)\ominus(\tyng{1,1},1)}_{(\phi^{(0)})^3\psi^{(0)}}$ \\\hline
\end{tabular}
\end{center}\vfill\eject
A similar table can be made for the magnetic theory\nopagebreak
\begin{center}
\begin{tabular}{|c|l|}
\hline
$3\Xi$ &  $SU(6)\times SU(2)_R$ \\\hline
$0$ & $(1,1)$ \\\hline
$2$ & $\underbrace{(\tyng{1,1},1)}_{\phi^{(0)}_M}$ \\\hline
$4$ & $\underbrace{(\tyng{2,2},1)\oplus(\tyng{1,1,1,1},1)}_{(\phi^{(0)}_M)^2}\ominus\underbrace{(\tyng{1,1,1,1},1)}_{\psi^{(0)}_M}$ \\\hline
$5$ & $\underbrace{(\tyng{1,1},\tyng{1})}_{\phi^{(1)}_M}$ \\\hline
$6$ & $\underbrace{(\tyng{3,3},1)\oplus(\tyng{2,2,1,1},1)\oplus(1,1)}_{(\phi^{(0)}_M)^3}\ominus\underbrace{(\tyng{2,2,1,1},1)\ominus(\tyng{2,1,1,1,1},1)\ominus(1,1)}_{\phi^{(0)}_M\psi^{(0)}_M}$ \\\hline
$7$ & $\underbrace{(\tyng{2,2},\tyng{1})\oplus(\tyng{2,1,1},\tyng{1})\oplus(\tyng{1,1,1,1},\tyng{1})}_{\phi^{(0)}_M\phi^{(1)}_M}\ominus\underbrace{(\tyng{1,1,1,1},\tyng{1})}_{\psi^{(1)}_M}$ \\\hline
$8$ & $\underbrace{(\tyng{4,4},1)\oplus(\tyng{3,3,1,1},1)\oplus(\tyng{2,2,2,2},1)\oplus(\tyng{1,1},1)}_{(\phi^{(0)}_M)^4}\oplus\underbrace{(\tyng{1,1},\tyng{2})}_{\phi^{(2)}_M}$ \\
& $\ominus\underbrace{(\tyng{3,3,1,1},1)\ominus(\tyng{3,2,1,1,1},1)\ominus(\tyng{1,1},1)\ominus(\tyng{2,2,2,2},1)\ominus(\tyng{2,2,2,1,1},1)\ominus(\tyng{1,1},1)}_{(\phi^{(0)}_M)^2\psi^{(0)}_M}\oplus\underbrace{(\tyng{2,2,2,1,1},1)}_{(\psi^{(0)}_M)^2}$ \\\hline
\end{tabular}
\end{center}
After taking into account cancellations between fermionic and bosonic states, those two expansions agree. From the physics of Seiberg duality it is expected, that the indices calculated on the electric and on the magnetic side actually agree to all orders in the expansion. This leads to the conjecture of a fairly involved group/number theory identity:
\beq\label{indeqsda}
\int\limits_{SU(2)}[dg]\exp\left(\sum\limits_{l=1}^\infty\frac{1}{l}f_e(t^l,\gamma^l,h^l,g^l)\right)=\exp\left(\sum\limits_{l=1}^\infty\frac{1}{l}f_m(t^l,\gamma^l,h^l)\right),
\eeq
where
\begin{align}
f_e(t,\gamma,h,g)&=\frac{(2t^2-t\chi_f(\gamma))\chi_3(g)+(t^{\frac{1}{3}}\chi_6(h)-t^{\frac{5}{3}}\chi_{\bar 6}(h))\chi_2(g)}{1-t\chi_f(\gamma)+t^2},\\
f_m(t,\gamma,h)&=\frac{t^{\frac{2}{3}}\chi_{15}(h)-t^{\frac{4}{3}}\chi_{\overline{15}}(h)}{1-t\chi_f(\gamma)+t^2}.
\end{align}

\subsection{More general Seiberg duality for SQCD in the conformal window}\label{secseibergb}

Calculating the superconformal index for more general $SU(N_c)$ SQCD with $N_f$ flavors and its Seiberg dual is quite cumbersome, but we will check a few interesting terms in the $t$ expansion. The 
oscillator modes contributing to the superconformal index are
\begin{center}
\begin{tabular}{|c|c|c|c|c|}
\hline
Field & $\Xi$ & $G$ & $H$ & $SU(2)_R$ \\\hline
$\phi^{(l)}_Q$ & $1-\frac{N_c}{N_f}+l$ & $f$ & $(N_f,1)_1$ & $l+1$ \\
$\phi^{(l)}_{\bar Q}$ & $1-\frac{N_c}{N_f}+l$ & $\bar f$ & $(1,\overline{N_f})_{-1}$ & $l+1$ \\
$\psi^{(l)}_Q$ & $1+\frac{N_c}{N_f}+l$ & $\bar f$ & $(\overline{N_f},1)_{-1}$ & $l+1$ \\
$\psi^{(l)}_{\bar Q}$ & $1+\frac{N_c}{N_f}+l$ & $f$ & $(1,N_f)_{1}$ & $l+1$ \\
$\lambda^{(l)}$ & $1+l$ & $Adj$ & $(1,1)_{0}$ & $l\oplus (l+2)$ \\
$F^{(l)}$ & $2+l$ & $Adj$ & $(1,1)_{0}$ & $(l+1)\oplus (l+1)$ \\\hline
\end{tabular}
\end{center}
for the electric theory, where $G=SU(N_c)$ and $H$ is the $SU(N_f)\times SU(N_f)\times U(1)_B$ flavor symmetry group, and
\begin{center}
\begin{tabular}{|c|c|c|c|c|}
\hline
Field & $\Xi$ & $G^\prime$ & $H$ & $SU(2)_R$ \\\hline
$\phi^{(l)}_q$ & $\frac{N_c}{N_f}+l$ & $f$ & $(\overline{N_f},1)_{\frac{N_c}{N_f-N_c}}$ & $l+1$ \\
$\phi^{(l)}_{\bar q}$ & $\frac{N_c}{N_f}+l$ & $\bar f$ & $(1,N_f)_{-\frac{N_c}{N_f-N_c}}$ & $l+1$ \\
$\phi^{(l)}_M$ & $2-2\frac{N_c}{N_f}+l$ & $1$ & $(N_f,\overline{N_f})_0$ & $l+1$ \\
$\psi^{(l)}_q$ & $2-\frac{N_c}{N_f}+l$ & $\bar f$ & $(N_f,1)_{-\frac{N_c}{N_f-N_c}}$ & $l+1$ \\
$\psi^{(l)}_{\bar q}$ & $2-\frac{N_c}{N_f}+l$ & $f$ & $(1,\overline{N_f})_{\frac{N_c}{N_f-N_c}}$ & $l+1$ \\
$\psi^{(l)}_M$ & $2\frac{N_c}{N_f}+l$ & $1$ & $(\overline{N_f},N_f)_0$ & $l+1$ \\
$\lambda^{(l)}$ & $1+l$ & $Adj$ & $(1,1)_{0}$ & $l\oplus (l+2)$ \\
$F^{(l)}$ & $2+l$ & $Adj$ & $(1,1)_{0}$ & $(l+1)\oplus (l+1)$ \\\hline
\end{tabular}
\end{center}
for the magnetic theory, with $G^\prime=SU(N_f-N_c)$.  We will now assume that $\frac{1}{3}\le\frac{N_c}{N_f}\le\frac{2}{3}$ is generic and $N_c$ is large.

In the electric theory there are the following contributions to the superconformal index
\begin{center}
\begin{tabular}{|c|l|}
\hline
$\Xi$ &  $H\times SU(2)_R$ \\\hline
$0$ & $(1,1,1)_0$ \\\hline
$2\frac{N_c}{N_f}$ & -- \\\hline
$1+2\frac{N_c}{N_f}$ & -- \\\hline
$2-2\frac{N_c}{N_f}$ & $\underbrace{(f,\bar f,1)_0}_{\phi^{(0)}_Q\phi^{(0)}_{\bar Q}}$ \\\hline
$2$ & $\ominus\underbrace{(1,1,1)_0\ominus(Adj,1,1)_0}_{\phi^{(0)}_Q\psi^{(0)}_Q}\ominus\underbrace{(1,1,1)_0\ominus(1,Adj,1)_0}_{\phi^{(0)}_{\bar Q}\psi^{(0)}_{\bar Q}}$ \\
& $\oplus\underbrace{(1,1,1)_0}_{(\lambda^{(0)})^2}$\\\hline
$2+2\frac{N_c}{N_f}$ & $\underbrace{(\bar f,f,1)_0}_{\psi^{(0)}_Q\psi^{(0)}_{\bar Q}}$ \\\hline
\end{tabular}
\end{center}
In the magnetic theory the contributions look quite different
\vfill\eject
\begin{center}
\begin{tabular}{|c|l|}
\hline
$\Xi$ &  $H\times SU(2)_R$ \\\hline
$0$ & $(1,1,1)_0$ \\\hline
$2\frac{N_c}{N_f}$ & $\underbrace{(\bar f,f,1)_0}_{\phi^{(0)}_q\phi^{(0)}_{\bar q}}\ominus\underbrace{(\bar f,f,1)_0}_{\psi^{(0)}_M}$ \\\hline
$1+2\frac{N_c}{N_f}$ & $\ominus\underbrace{(\bar f,f,2)_0}_{\phi^{(0)}_q\phi^{(0)}_{\bar q}\lambda^{(0)}}\oplus\underbrace{(\bar f,f,2)_0}_{\phi^{(0)}_q\phi^{(1)}_{\bar q}}\oplus\underbrace{(\bar f,f,2)_0}_{\phi^{(1)}_q\phi^{(0)}_{\bar q}}\ominus\underbrace{(\bar f,f,2)_0}_{\psi^{(1)}_M}$\\\hline
$2-2\frac{N_c}{N_f}$ & $\underbrace{(f,\bar f,1)_0}_{\phi^{(0)}_M}$ \\\hline
$2$ & $\ominus\underbrace{(1,1,1)_0\ominus(Adj,1,1)_0}_{\phi^{(0)}_q\psi^{(0)}_q}\ominus\underbrace{(1,1,1)_0\ominus(1,Adj,1)_0}_{\phi^{(0)}_{\bar q}\psi^{(0)}_{\bar q}}$\\
& $\ominus\underbrace{(1,1,1)_0\ominus(Adj,1,1)_0\ominus(1,Adj,1)_0\ominus(Adj,Adj,1)_0}_{\phi^{(0)}_M\psi^{(0)}_M}$\\
& $\oplus\underbrace{(1,1,1)_0\oplus(Adj,1,1)_0\oplus(1,Adj,1)_0\oplus(Adj,Adj,1)_0}_{\phi^{(0)}_q\phi^{(0)}_{\bar q}\psi^{(0)}_M}\oplus\underbrace{(1,1,1)_0}_{(\lambda^{(0)})^2}$\\\hline
$2+2\frac{N_c}{N_f}$ & $\underbrace{(f\otimes\Lambda^2\bar f,\bar f\otimes S^2f,1)_0\oplus (f\otimes S^2\bar f,\bar f\otimes\Lambda^2f,1)_0}_{\phi^{(0)}_M(\psi^{(0)}_M)^2}$ \\
& $\oplus\underbrace{(f\otimes\Lambda^2\bar f,\bar f\otimes\Lambda^2f,1)_0\oplus (f\otimes S^2\bar f,\bar f\otimes S^2f,1)_0}_{\phi^{(0)}_M(\phi^{(0)}_q)^2(\phi^{(0)}_{\bar q})^2}$ \\
& $\ominus\underbrace{(f\otimes\Lambda^2\bar f,\bar f\otimes\Lambda^2f,1)_0\ominus (f\otimes\Lambda^2\bar f,\bar f\otimes S^2f,1)_0}_{\phi^{(0)}_M\psi^{(0)}_M\phi^{(0)}_q\phi^{(0)}_{\bar q}}$ \\
& $\ominus\underbrace{(f\otimes S^2\bar f,\bar f\otimes\Lambda^2f,1)_0\ominus (f\otimes S^2\bar f,\bar f\otimes S^2f,1)_0}_{\phi^{(0)}_M\psi^{(0)}_M\phi^{(0)}_q\phi^{(0)}_{\bar q}}$ \\
& $\ominus\underbrace{(f\otimes\Lambda^2\bar f,f,1)_0\ominus (f\otimes S^2\bar f,f,1)_0}_{\psi^{(0)}_q(\phi^{(0)}_q)^2\phi^{(0)}_{\bar q}}\ominus\underbrace{(\bar f,\bar f\otimes\Lambda^2f,1)_0\ominus (\bar f,\bar f\otimes S^2f,1)_0}_{\psi^{(0)}_{\bar q}(\phi^{(0)}_{\bar q})^2\phi^{(0)}_q}$ \\
& $\oplus\underbrace{(f\otimes\Lambda^2\bar f,f,1)_0\oplus (f\otimes S^2\bar f,f,1)_0}_{\psi^{(0)}_q\phi^{(0)}_q\psi^{(0)}_M}\oplus\underbrace{(\bar f,\bar f\otimes\Lambda^2f,1)_0\oplus (\bar f,\bar f\otimes S^2f,1)_0}_{\psi^{(0)}_{\bar q}\phi^{(0)}_{\bar q}\psi^{(0)}_M}$ \\
& $\oplus\underbrace{(\bar f,f,1)_0\oplus (\bar f,f,1)_0}_{F^{(0)}\phi^{(0)}_q\phi^{(0)}_{\bar q}}\ominus\underbrace{(\bar f,f,1)_0}_{(\lambda^{(0)})^2\psi^{(0)}_M}\ominus\underbrace{(\bar f,f,1)_0\ominus (\bar f,f,3)_0}_{\lambda^{(0)}\phi^{(1)}_q\phi^{(0)}_{\bar q}}$ \\
& $\ominus\underbrace{(\bar f,f,1)_0\ominus (\bar f,f,3)_0}_{\lambda^{(0)}\phi^{(0)}_q\phi^{(1)}_{\bar q}}\ominus\underbrace{(\bar f,f,1)_0\ominus (\bar f,f,3)_0}_{\lambda^{(1)}\phi^{(0)}_q\phi^{(0)}_{\bar q}}\ominus\underbrace{(\bar f,f,3)_0}_{\psi^{(2)}_M}\oplus\underbrace{(\bar f,f,3)_0}_{\phi^{(0)}_q\phi^{(2)}_{\bar q}}$\\
& $\oplus\underbrace{(\bar f,f,3)_0}_{\phi^{(2)}_q\phi^{(0)}_{\bar q}}\oplus\underbrace{(\bar f,f,1)_0\oplus\bar f,f,3)_0}_{\phi^{(1)}_q\phi^{(1)}_{\bar q}}\oplus\underbrace{(\bar f,f,1)_0\oplus(\bar f,f,1)_0\oplus\bar f,f,3)_0}_{(\lambda^{(0)})^2\phi^{(0)}_q\phi^{(0)}_{\bar q}}$\\\hline
\end{tabular}
\end{center}
but there are cancellations such that the conformal indices of the electric and the magnetic theories agree at the given levels. Note that we checked only states with vanishing baryon number, as is usual for large $N_c$. This again leads to the conjecture of a fairly involved group/number theory identity:
\begin{multline}\label{indeqsdb}
\int\limits_{SU(N_c)}[dg_1]\exp\left(\sum\limits_{l=1}^\infty\frac{1}{l}f_e(t^l,\gamma^l,h_1^l,h_2^l,b^l,g_1^l)\right)\\
=\int\limits_{SU(N_f-N_c)}[dg_2]\exp\left(\sum\limits_{l=1}^\infty\frac{1}{l}f_m(t^l,\gamma^l,h_1^l,h_2^l,b^l,g_2^l)\right)
\end{multline}
where $f_e$ and $f_m$ are given by (\ref{genletters}) using the matter content of the electric and the magnetic theory respectively and $(h_1,h_2)_b\in SU(N_f)\times SU(N_f)\times U(1)_B$.

At the lowest orders in the $t$-expansion there seem to be mainly contributions from chiral operators and some 'relations' in the form of fermions. However, at higher orders the structure is much more complicated. Also, the gauginos typically contribute at levels, where a lot of other modes contribute. For this reason it seems hard to separate out the gaugino contributions.

\section{Discussion}

We have established, that the superconformal index can be calculated by reducing a interacting field theory at the conformal fixed point to a free $\CN=1$ supersymmetric field theory on $S^3\times\BR$. Using this result we checked a few examples of field theory dualities and found agreement at lower orders in the $t$-expansion of the full superconformal index.

It would be very interesting to prove in general that the superconformal index of dual theories agrees. The simple dependence of equations (\ref{genletters}) and (\ref{pexpind}) for the superconformal index on the field content and the universal form of the denominator of $f$ suggests, that there might be a simple geometrical argument involving the moduli space of vacua and the chiral ring or anomaly matching.

There are also still interesting examples involving String theory on $AdS_5\times Y^{(p,q)}$ \cite{Gauntlett:2004yd,Martelli:2004wu} and its CFT duals \cite{Benvenuti:2004dy}. In many of those examples the field theory descriptions are related by Seiberg dualities. Those models might be treated using large $N$ matrix model technologies \cite{Kinney:2005ej,Nakayama:2005mf}.

Also, it is clear from the tables in sections \ref{secseiberga} and \ref{secseibergb} that there are net contributions to the superconformal index from fermions and from bosons coming in different representations of the global symmetry group. This might partly explain the missing states in the black hole entropy counting of \cite{Kinney:2005ej}. Those different states cannot really be seen very well by just looking at the characters without decomposing them into irreps: If the characters are written in terms of Cartan generators, often the coefficients of those have all the same sign, but when decomposed into irreps there are contributions with different signs.

It would be interesting to do an argument similar to the spectral sequence argument in section \ref{ssecwz} to see whether the superconformal index actually counts all the $\frac{1}{4}$ BPS states in a generic theory or whether there are bigger cancellations inside the superconformal index of an interacting theory. This might lead to a description of those $\frac{1}{4}$ BPS states similar to \cite{Berenstein:2004kk} but with much less supersymmetry. This might be compared to gravitational black hole states or D-brane configurations similar to \cite{Mikhailov:2000ya,Beasley:2002xv,Mandal:2006tk,Biswas:2006tj}.

%The correspondence between gravitational configurations and certain BPS states in the field theory has been explored in quite some detail \cite{Berenstein:2004kk,Lin:2004nb}. However, this analysis is using a lot of supersymmetry. 

\begin{appendix}

\section{Some conventions}

We follow the Metric conventions $(-+++)$ and the gamma matrices are
\begin{align}
\gamma^0&=\left(\begin{array}{cc} 0 & -\Bid \\ \Bid & 0 \end{array}\right), 
\quad
\gamma^i=2\left(\begin{array}{cc} 0 & \sigma^i \\ \sigma^i & 0 \end{array}\right),\\
\gamma^5&=i\gamma^{0123}=
\left(\begin{array}{cc} \Bid & 0 \\ 0 & -\Bid \end{array}\right),
\end{align}
where the Pauli matrices are
\beq
\sigma_{(P)}^1=
\half\left(\begin{array}{cc} 0 & 1 \\ 1 & 0 \end{array}\right), \quad
\sigma_{(P)}^2=
\half\left(\begin{array}{cc} 0 & -i \\ i & 0 \end{array}\right), \quad
\sigma_{(P)}^3=
\half\left(\begin{array}{cc} 1 & 0 \\ 0 & -1 \end{array}\right).
\eeq

The Hermitean conjugation is defined as
\beq
(\gamma^\mu)^\dagger=-C\gamma^\mu C^{-1}\quad\and\quad\bar\psi=\psi^\dagger C\quad\mathrm{with}\quad C=\gamma^0,
\eeq
the transpose as
\beq
(\gamma^\mu)^t=D\gamma^\mu D^{-1}\quad\and\quad\tilde\psi=\psi^t D\quad\mathrm{with}\quad D=\gamma^{135}
\eeq
and the complex conjugation as
\beq
(\gamma^\mu)^\ast=-B\gamma^\mu B^{-1}\quad\and\quad\ast\psi=B^{-1}\psi^\ast\quad\mathrm{with}\quad B=\gamma^2.
\eeq

\section{Lagrangian for a $\CN=1$ gauge theory on $S^3\times\BR$}\label{applagr}

The Lagrangian for a $\CN=1$ gauge theory on $S^3\times\BR$ \cite{Romelsberger:2005eg} is written down in a vielbein basis of right invariant 1-forms and the spatial derivatives are derivatives with respect to the Killing vector fields generated by $SU(2)_L$.
\begin{align}
\CL_g&=\frac{1}{g^2}\bigg(4\tr\CF_{0i}\CF_{0i}-8\tr\CF_{ij}\CF_{ij}\\\nonumber
&+i\tr\bar\lambda\gamma^0\CD_0\lambda-
2i\tr\bar\lambda\gamma^i
\left(\CD_i+\frac{1}{8}\epsilon_{ijk}\gamma^{jk}\right)\lambda-\tr D^2\bigg),\\
\CL_\theta&=\theta\epsilon_{ijk}\CF_{0i}\CF_{jk},\\
%\CL_{FI}&=\kappa\,\tr(D-\CA_0),\\
\CL_0&=\left(\CD_0-i\frac{3q-2}{2}\right)\phi^\dagger
\left(\CD_0+i\frac{3iq-2}{2}\right)\phi-
4\CD_i\phi^\dagger\CD_i\phi-\phi^\dagger\phi\\\nonumber
&+i\bar\psi\gamma^0\left(\CD_0+i\frac{3q-2}{2}\right)\psi-
2i\bar\psi\gamma^i\left(\CD_i+\frac{1}{8}\epsilon_{ijk}\gamma^{jk}\right)\psi+
F^\dagger F\\\nonumber
&+2i\phi^\dagger D^{(\rho)}\phi-
2i\phi^\dagger\tilde\lambda^{(\rho)}\psi+
2\bar\psi(\ast\lambda)^{(\rho)}\phi,\\
\CL_W&=W^\prime(\phi)F-\half W^{\prime\prime}(\phi)\tilde\psi\psi+{\rm h.c.}
\end{align}

\end{appendix}

%\nopagebreak
\section*{Acknowledgments}

I would like to thank F.~Dolan, I.~Egusquiza, A.~Hanany, Y.~He, J.~Ma\~nes, W.~Nahm, C.~S\"amann and M.~Valle for very fruitful discussions. I would like to thank the Departamento de Fisica Teorica e Historia de la Ciencia of the Universidad del Pais Vasco and the Benasque center for science for their hospitality.

%\end{acknowledgments}

\providecommand{\href}[2]{#2}\begingroup\raggedright\endgroup

%\bibliographystyle{my-h-elsevier}
%\bibliographystyle{utphys}
%\bibliography{index2}

%\begin{thebibliography}{10}

% bibliography entries should follow the format of the sample items
% below, without the "%" at the beginning of the line.

%\bibitem{vgeemen}
%A. Albano and S. Katz,
%Van Geemen's Families of Lines on Special
%Quintic Threefolds,
%Manuscripta Math. {\bf 70} (1991) 183.

%\bibitem{AK}
%A. Albano and S. Katz,
%Lines on The Fermat Quintic Threefold,
%Trans. AMS. {\bf 324} (1991) 353.

%\bibitem{ADD}
%S.K. Ashok, E. Dell'Aquila and D.-E. Diaconescu,
%Fractional Branes in Landau-Ginzburg Orbifolds,
%hep-th/0401135.

%\end{thebibliography}

\end{document}